\documentclass{jfm}
\usepackage{amsmath}
\usepackage{amsfonts}
\usepackage{graphicx}
\usepackage{chemarr, amssymb}
\usepackage{color}
\usepackage{subfigure}
\usepackage{natbib}

 \usepackage{placeins}

\newcommand{\bc}{\ensuremath{\mathbf{c}}}

\newcommand{\bx}{\ensuremath{\mathbf{x}}}

\newcommand{\Kn}{\ensuremath{{\rm Kn}}}

\title[Hydrodynamics in a micro-cavity]
{Exploring Hydrodynamic Closures for the Lid-driven Micro-cavity}

\author[S. Ansumali, C. E. Frouzakis, I. V. Karlin and I. G.
Kevrekidis] {S. Ansumali$^1$, C. E. Frouzakis$^1$,  I. V.
Karlin$^1$\\ and I. G. Kevrekidis$^2$}
\affiliation{$^1$Aerothermochemistry and Combustion Systems
Laboratory, Swiss Federal Institute of Technology Zurich (ETHZ),
8092  Z\"urich, Switzlerland \\ [\affilskip] $^2$Department of
Chemical Engineering, Princeton University, NJ 08544-5263, USA }

\begin{document}

\maketitle

\begin{abstract}
A minimal kinetic model is used to study analytically and
numerically flows at a micrometer scale. Using the lid-driven
microcavity as an illustrative example, the interplay between
kinetics and hydrodynamics is quantitatively visualized. The
validity of various theories of non-equilibrium thermodynamics of
flowing systems is tested in this nontrivial microflow.

\end{abstract}

\section{Introduction}

Flows in microdevices constitute an emerging application field of
fluid dynamics \cite[][]{Karniadakis2}. Despite impressive
experimental progress, theoretical understanding of microflows
remains incomplete. For example, even though microflows are highly
subsonic, the assumption of incompressible fluid motion is often
questionable (see, e. g. \cite{Garcia}). Interactions between the
relaxation of density variations, the rarefaction and the flow
geometry are not completely understood. One of the reasons for
this is the lack of commonly accepted  models for efficient
simulations of microflows.

In principle, microflows can be studied using molecular level
methods, such as the Direct Simulation Monte Carlo (DSMC) method
\cite[]{bird}. However, molecular dynamics methods face severe
limitations for subsonic flows. The number of particles required
for simulations in realistic geometries with large aspect ratios
and the number of time steps needed to reach the statistical
steady state are prohibitively high \cite[]{DSMC} (the time step
of DSMC is $\sim 10^{-10}$sec, while relevant physical time-scales
are $\sim 10-100$sec). Moreover, a detailed microscopic
description in terms of the particle distribution function is not
necessary for design/endineering purposes. Thus, development of
reduced models enabling efficient simulations is an important
issue.

In this paper we consider a two-dimensional  minimal kinetic model
with nine discrete velocities \cite[]{DHT,AK5}. It has been
recently shown by several groups
\cite[]{AK4,AK2004a,ELBMMICRO,SUCCIDBC} that this model compares
well with analytical results of kinetic theory \cite[]{Cerci} in
simple flow geometries (channel flows), as well as with molecular
dynamics simulations. The focus of this paper is the validation of
several theoretical models of (extended) hydrodynamics versus
direct numerical simulation in a non-trivial flow. For that
purpose, we considered the flow in a lid-driven cavity. We will
visualize the onset of the hydrodynamic description, the effect of
the boundaries etc.

The paper is organized as follows: For completeness, the kinetic
model \cite[]{DHT,AK5} is briefly presented in section
\ref{secII}. In section \ref{secIII}, we show the relation of our
model to the well-known Grad moment system derived from the
Boltzmann kinetic equation \cite[]{GradH}. We compare analytically
the dispersion relation for the present model and the Grad moment
system. This comparison reveals that the kinetic model of section
\ref{secII} is a superset of Grad's moment system. In section
\ref{secIV}, a parametric numerical study of the flow in a
micro-cavity is presented. Results are also compared to a DSMC
simulation. In section \ref{secV}, the reduced description of the
model kinetic equation is investigated, and a visual
representation quantifying the onset and breakdown of the
hydrodynamics is discussed. We conclude in section \ref{secVI}
with a spectral analysis of the steady state flow and some
suggestions for further research
\cite[]{Yannis3,Yanis1,Yanis,Yannis4}.

\section{\label{secII} Minimal kinetic model}

We consider the discrete velocity model with the following set of
nine discrete velocities:
\begin{align}
\label{dv}
\begin{split}
 c_x &= \left[ 0,1,0,-1,0,1-1,-1,1\right],\quad c_y =
\left[0,0,1,0,-1,1,1,-1,-1\right].
\end{split}
\end{align}
The local hydrodynamic fields are defined in terms of the discrete
population, $f_i$, as:
\begin{equation}
\label{hfields}
 \sum_{i=1}^{9} f_i  \{ 1,\, c_{x\,i}, \, c_{y\,i}\}=\{\rho,\, j_x, \, j_y\},
\end{equation}
where $\rho$ is the local mass density, and $j_{\alpha}$ is the
local momentum density of the model. The populations $f_i \equiv
f(\bx,\bc_{i},t)$ are functions of the discrete velocity
$\bc_{i}$, position $\bx$ and time $t$. We consider the following
kinetic equation for the populations (the Bhatnagar-Gross-Krook
single relaxation time model):
\begin{equation}
\label{LBM}
\partial_t f_i+ \bc_{i} \cdot \partial_{\bx}f_i =
-\frac{1}{\tau} \left(f_i- f_i^{\rm eq} (f) \right),
\end{equation}
where $\tau$ is the relaxation time, and $f_i^{\rm eq}$ is the
local equilibrium \cite[]{AK5}:
\begin{align}
\label{TED}
\begin{split}
  f^{\rm eq}_i =\rho W_i
  \left(2 -\sqrt{1+ 3 {u_{x}}^2}\right)\left(2 -
    \sqrt{1+3{u_{y}}^2}\right)\left(\frac{ 2 u_{x}+
    \sqrt{1+3{u_{x}}^2}}{1- u_x} \right)^{c_{ x i}}\left(\frac{2 u_{y} +
    \sqrt{1+ 3{u_{y}}^2}}{1-u_y} \right)^{c_{ y i}},
\end{split}
\end{align}
where $u_{\alpha}= j_{\alpha}/\rho$, and the speed of sound is
$c_{\rm s}=1/\sqrt{3}$. The local equilibrium distribution
$f_i^{\rm eq}$ is the minimizer of the discrete $H$ function
\cite[]{DHT, AK5}:
\begin{equation}
\label{app:H} H=\sum_{i=1}^{9}
  f_{i}\ln\left(\frac{f_{i}}{W_i} \right), \; \mbox{with weights}
 \quad W = \left[\frac{16}{36}, \frac{4}{36}, \frac{4}{36}, \frac{4}{36},
                   \frac{4}{36}, \frac{1}{36}, \frac{1}{36}, \frac{1}{36},
                   \frac{1}{36} \right],
\end{equation}
under the constraints of the local hydrodynamic fields
(\ref{hfields}). Note the important factorization over spatial
components of the equilibrium (\ref{TED}). This is similar to the
familiar property of the local Maxwell distribution, and it
distinguishes (\ref{TED}) among other discrete-velocity
equilibria.

In the hydrodynamic regime, the model recovers the Navier-Stokes
equation with viscosity coefficient  $\mu = p \tau$, where $p=
\rho c_{\rm s}^2$ is the pressure (see section \ref{secV}). The
diffusive-wall approximation \cite[]{AK4} is used for wall
boundary conditions.

\section{\label{secIII} Grad's moment system and the minimal kinetic model }

\subsection{The moment system}

It is  useful to represent the discrete velocity model (\ref{LBM})
in the form of a moment system. For simplicity, we shall consider
the linearized version of the model. Note that linearization is
required only for the collision term on the right hand side of
equation (\ref{LBM}). This is at variance with Grad's moment
systems \cite[]{GradH} where the advection terms are nonlinear as
well. We choose the following nine non-dimensional moments as
independent variables:

\begin{align}
\label{moms} {M} =  \left[\frac{     \rho}{ \rho_0}, \frac{
j_x}{\rho_0 c_{\rm s}},
                       \frac{ j_y}{\rho_0 c_{\rm s}}, \frac{ P}{\rho_0 c_{\rm s}^2} ,
                       \frac{ N}{\rho_0 c_{\rm s}^2}, \frac{P_{xy}}{\rho_0 c_{\rm s}^2},
                       \frac{ q_{x}}{2 \rho_0 c_{\rm s}^3}, \frac{ q_{y}}{2 \rho_0 c_{\rm s}^3},
                       \frac{  \psi}{2 \rho_0 c_{\rm s}^4} \right],
\end{align}

where
 \begin{equation}
 \psi =R_{yyyy}+R_{xxxx}- 2 R_{xxyy},
 \end{equation}
is a scalar obtained from 4$^{th}$-order moments $R_{\alpha \beta
\gamma \theta}=\sum_{i=1}^{9}  f_i c_{\alpha i} c_{ \beta i } c_{
\gamma i} c_{ \theta i}$, ${ N}=\sum_{i=1}^{9}  f_i (c_{xi }^2
-c_{yi }^2)/2\equiv {(P_{xx}- P_{y y})}/{2}$ is the difference of
the normal stresses, $P=\sum_{i=1}^{9} f_i c_i^2$ is the trace of
the pressure tensor, and $q_{\alpha} =\sum_{i=1}^{9} f_i c_{
\alpha i} c_{i}^2 $ is the energy flux obtained by contraction of
the third-order moment $Q_{\alpha \beta \gamma}=\sum_{i=1}^{9} f_i
c_{ \alpha i} c_{ \beta i} c_{ \gamma i}$. Time and space are made
non-dimensional in such a way that for a fixed system size $L$
they are measured in the units of mean free time and mean free
path, respectively: $\bx^{\prime}= \bx /(L \Kn), t^{\prime} = t/
\tau$, where $\Kn= \tau c_{\rm s} / L$ is the Knudsen number. The
linearized equations for the moments $M$ (\ref{moms}) read (from
now on we  use the same notation for the non-dimensional
variables):

\begin{align}
 \label{LinMom}
 \begin{split}
 \partial_t \rho +  \, \partial_x j_x +   \partial_y j_y &= 0, \\
 \partial_t j_x + \partial_x  \left(P+ N \right)+  \partial_y P_{x y} &= 0, \\
 \partial_t j_y +  \partial_x P_{xy} + \partial_y \left(  P- N \right) &= 0,\\
  \partial_t P + \partial_x q_x + \partial_y q_y  &=\left({\rho}- P\right), \\
 \partial_t N +  \partial_x\left(q_x - Q_{xyy} \right) -   \partial_y
    \left( q_y - Q_{yxx} \right)  &= -N,\\
 \partial_t P_{xy} +   \partial_x Q_{y xx} +    \partial_y Q_{y yx} &= -P_{xy},\\
 \partial_t q_x +   \partial_x R_{xx\alpha \alpha} +   \partial_y R_{xy \alpha\alpha} &=
    \left( 2   j_x - q_x\right), \\
 \partial_t q_y +  \partial_x R_{xy\alpha \alpha} +  \partial_y R_{y y\alpha\alpha} &=
    \left( 2  j_y - q_y\right),\\
 \partial_t \psi +  \partial_x \left( j_x -  q_x\right) +  \partial_y\left( j_y - q_y \right) &=
    \left(2 \rho - \psi\right).
\end{split}
\end{align}
Furthermore, by construction of the discrete velocities
(\ref{dv}), the following relations are satisfied:
\begin{align}
 \label{LinMomL}
 Q_{xyy} = 2 q_x - {3}  j_x, \quad
 Q_{yxx} =  2 q_y  -  {3} j_y, \\
  \label{LinMomLOne}
 R_{x y \alpha \alpha } = 3 P_{xy},  \; \;
 R_{x x \alpha \alpha} = 3 \left(P+  \frac{1}{2} N\right) - \frac{1}{2} \psi,  \; \;
 R_{yy \alpha \alpha} = 3 \left(P- \frac{1}{2} N\right)- \frac{1}{2} \psi.
\end{align}
Apart from the lack of conservation of the energy and linearity of
the advection, equation (\ref{LinMom}) is quite similar to Grad's
two-dimensional $8$-moment system \footnote{The variables used in
the $D$-dimensional Grad's system are density, $D$ components of
the momentum flux, $D(D+1)/2$ components of the pressure tensor
and $D$ components of the energy flux. The number of fields in
Grad's system is $8$ for $D=2$ and $13$ for $D=3$.}. However, in
the present case a particular component of the 4$^{th}$-order
moment is also included as a variable. In other words, Grad's
non-linear closure for the 4$^{th}$-order moment is replaced by an
evolution equation with a linear advection term. We note here that
while the formulation of boundary conditions for Grad's moment
system remains an open problem, the boundary conditions for the
extended moment system (system \ref{LinMom}, \ref{LinMomL}, and
\ref{LinMomLOne}) are well established through its
discrete-velocity representation (\ref{LBM}) \cite[]{AK4}. We also
note that like any other Grad's system the present model
reproduces the Navier-Stokes equation in the hydrodynamic limit
\cite[]{DHT,AK5}. The moment system (\ref{LinMom}) reveals the
meaning of the densities appearing in model: The dimensionless
density is the dimensionless pressure of the real fluid in the low
Mach number limit, while the momentum flux density should be
identified with the velocity in the incompressible limit. With
this identification, we shall compare the moment system
(\ref{LinMom}) with Grad's system.

\subsection{One-dimensional Grad's moment system}

Since energy is not conserved by the model (\ref{LBM}), the
comparison will be with another Grad moment system which (for
$D=3$) is usually referred to as the $10$-moment system
\footnote{The variables used in this $D$-dimensional Grad's system
are density, $D$ components of the momentum flux, and $D(D+1)/2$
components of the pressure tensor, resulting in 6 and 10 variables
for $D=2$ and $D=3$, respectively.}. For one-dimensional flows,
the linearized  Grad's $10$-moment system can be written as
\cite[]{GKbook,KG203}:
\begin{align}
\label{Grad10}
\begin{split}
 \partial_t p + \gamma \partial_x u_x=0,\quad
 \partial_t u_x + \partial_x P_{xx} =0,\quad
 \partial_t P_{xx} + 3\partial_x u_x= - \left(P_{xx} - p\right),
\end{split}
\end{align}
where $\gamma$ is the ratio of the specific heats of the fluid,
and $\gamma=(D+2)/D$ for a $D$-dimensional dilute gas. This model
can be described in terms of its dispersion relation, which upon
substitution of the solution in the form $\sim \exp{\left(\omega t
+ i k x\right)}$ reads:
\begin{align}
\label{DR}
  \omega^3 + \omega^2 + 3 k^2 \omega + \gamma  k^2 = 0.
\end{align}
The low wave-number asymptotic represents the large-scale dynamics
(hydrodynamic scale of $\Kn\ll 1$), while  the high-wave number
limit represent the molecular scales quantified by $\Kn\gg 1$. The
low wave number ($\Kn\ll 1$) asymptotic, $\omega_{\rm l}$, and the
large wave number ($\Kn \gg 1$) asymptotic, $\omega_{\rm h}$, are:
 \begin{align}
  \omega_{\rm l} =\left\{\frac{(-3 + \gamma)}{2} k^2 \pm i \sqrt{\gamma} k , \;
      -1  -( -3 + \gamma ) k^2 \right\}, \; \;
  \omega_{\rm h}=\left\{\frac{(-3 + \gamma)}{6}\pm i \sqrt{3} k, \;
     -\frac{\gamma }{3} \right\}.
 \end{align}
The two complex conjugate modes (acoustic modes) of the $O(k^2)$
dynamics, are given by the first two roots of $\omega_{\rm l}$,
and represent the  hydrodynamic limit (the Navier-Stokes
approximation) of the model. The third root in this limit is real
and negative, corresponding to the relaxational behavior of the
non-hydrodynamic variable (stress): the dominant contribution
(equal to $-1$) is the relaxation rate towards the equilibrium
value, while the next-order correction suggests slaving of viscous
forces, which amounts to the constitutive relation for stress
($(-3 + \gamma)/2 k^2$ ). Furthermore, the $k^2$ dependence of the
relaxation term justifies the assumption of scale separation (the
higher the wave-number, the faster the relaxation). The real part
of the high wave-number solution $\omega_{\rm h}$ is independent
of $k$, which shows that the relaxation at very high Knudsen
number is the same for all wavenumbers (so-called ``Rosenau
saturation" \cite[]{GK96a,Slemrod98}). Thus, the assumption of
scale-separation is not valid for high Knudsen number dynamics.

\subsection{Dispersion relation for the moment system}

\begin{figure}
\begin{center}
  \includegraphics[width=6cm]{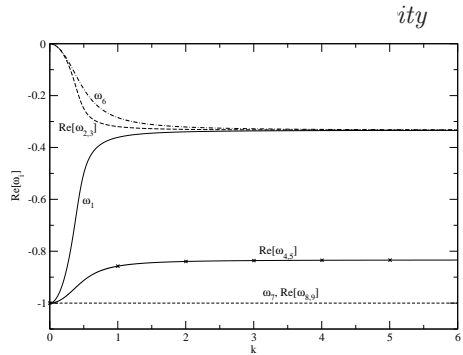}
  \caption{Real part of the solutions of the dispersion relation
  (equation (\ref{d2q9disp})).
  Roots $\omega_{2,3}$ and $\omega_1$ correspond to Grad's subsystem
  (equation \eqref{Grad10}). The real-valued root $\omega_6$
  and the complex conjugate roots $\omega_{2,3}$ are extended hydrodynamic
  modes.}\label{disp}
  \end{center}
\end{figure}

The dispersion relation for the one-dimensional version of the
moment system \eqref{LinMom} (i.e. neglecting all derivatives in
the $y$-direction) reads:
\begin{equation} \label{d2q9disp}
  (\omega^3 + \omega^2 + 3k^2  \omega + k^2)
  (\omega^3 + 2\omega^2 + (3k^2+1)\omega + k^2)
  (1+\omega)
  ((1+\omega^2)+2k^2) =0.
\end{equation}
The real parts of the roots of this equation (attenuation rates
${\rm Re} [\omega(k)]$) are plotted in Fig.\ \ref{disp} as
functions of the wave vector $k$.
It is clear that for one-dimensional flows, the dynamics of three
of the moments ($\rho$, $j_x$, and $P$) are decoupled from the
rest of the variables, and follows  of the dynamics of the
one-dimensional Grad's moment system (\ref{Grad10}) with
$\gamma=1$.

The similarity between Grad's moment system and the present model
is an important fingerprint of the kinetic nature of the latter
\footnote{Grad \cite[]{GradH} already mentioned that moment
systems are particularly well suited for low Mach number flows.
Qualitatively, this is explained as follows: when expansion in the
Mach number around the no-flow state is addressed, the first
nonlinear terms in the advection are of order $u^2/c_{\rm s}^2\sim
{\rm Ma}^2$. On the other hand, the same order in ${\rm Ma}$ terms
in the relaxation contribute $u^2/(\tau c_{\rm s}^2)\sim {\rm
Ma}^2/{\rm Kn}$. Thus, if Knudsen number is also small, nonlinear
terms in the advection can be neglected while the nonlinearity in
the relaxation should be kept. That is why the model (\ref{LBM}) -
linear in the advection and nonlinear in the relaxation - belongs
to the same domain of validity as Grad's moment systems for
subsonic flows.}. Note that in the case of two-dimensional flows,
the agreement between the present model and Grad's system is only
qualitative. The present moment system is isotropic only up to
$O(k^2)$. Thus, the dispersion relation of the model (\ref{LBM})
is expected to match the one of Grad's system only up to the same
order. In the hydrodynamic and slip-flow regime addressed below,
this order of isotropy is sufficient.
In the presence of boundaries and/or non-linearities, it is
necessary to resort to numerics. Below we use the entropic lattice
Boltzmann discretization method (ELBM) of the model (\ref{LBM}).

\section{ \label{secIV}Flow in a lid-driven micro-cavity}

The two-dimensional flow in a lid-driven cavity was simulated with
ELBM over a range of Knudsen numbers defined as ${\rm Kn}={\rm
Ma}/\rm Re$. In the simulations, the Mach number was fixed at
${\rm Ma}=0.01$ and the Reynolds number, $\rm Re$, was varied.
Initially, the fluid in the cavity is at rest and the upper wall
of the domain is impulsively set to motion with $u_{\rm
lid}=c_{\rm s}{\rm Ma}$. Diffusive boundary conditions are imposed
at the walls \cite[]{AK4}, and the domain was discretized using
$151$ points in each spatial direction. Time integration is
continued till the steady state is reached; matrix-free,
Newton-Krylov fixed point algorithms for the accelerated
computation of the steady state are also being explored
\cite[]{Yannis3}.

\subsection{Validation with DSMC simulation of the micro-cavity}

\begin{figure}
\begin{center}
\hbox{ \centerline{
  \includegraphics[width=4.0cm]{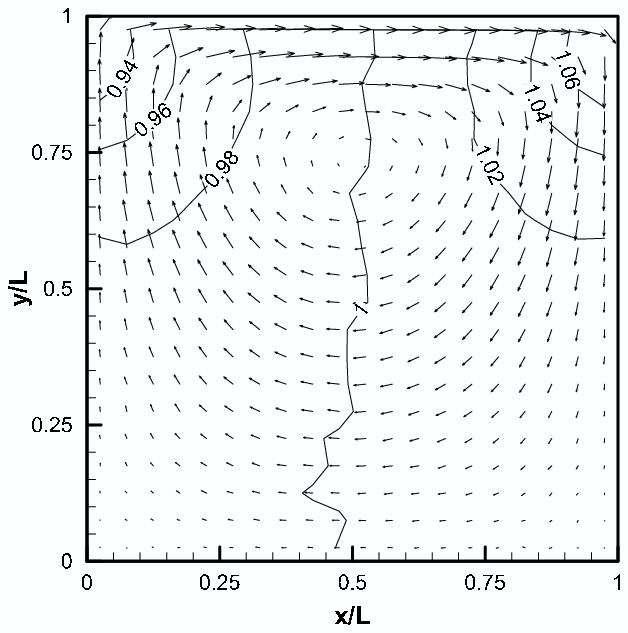}
  \includegraphics[width=3.9cm]{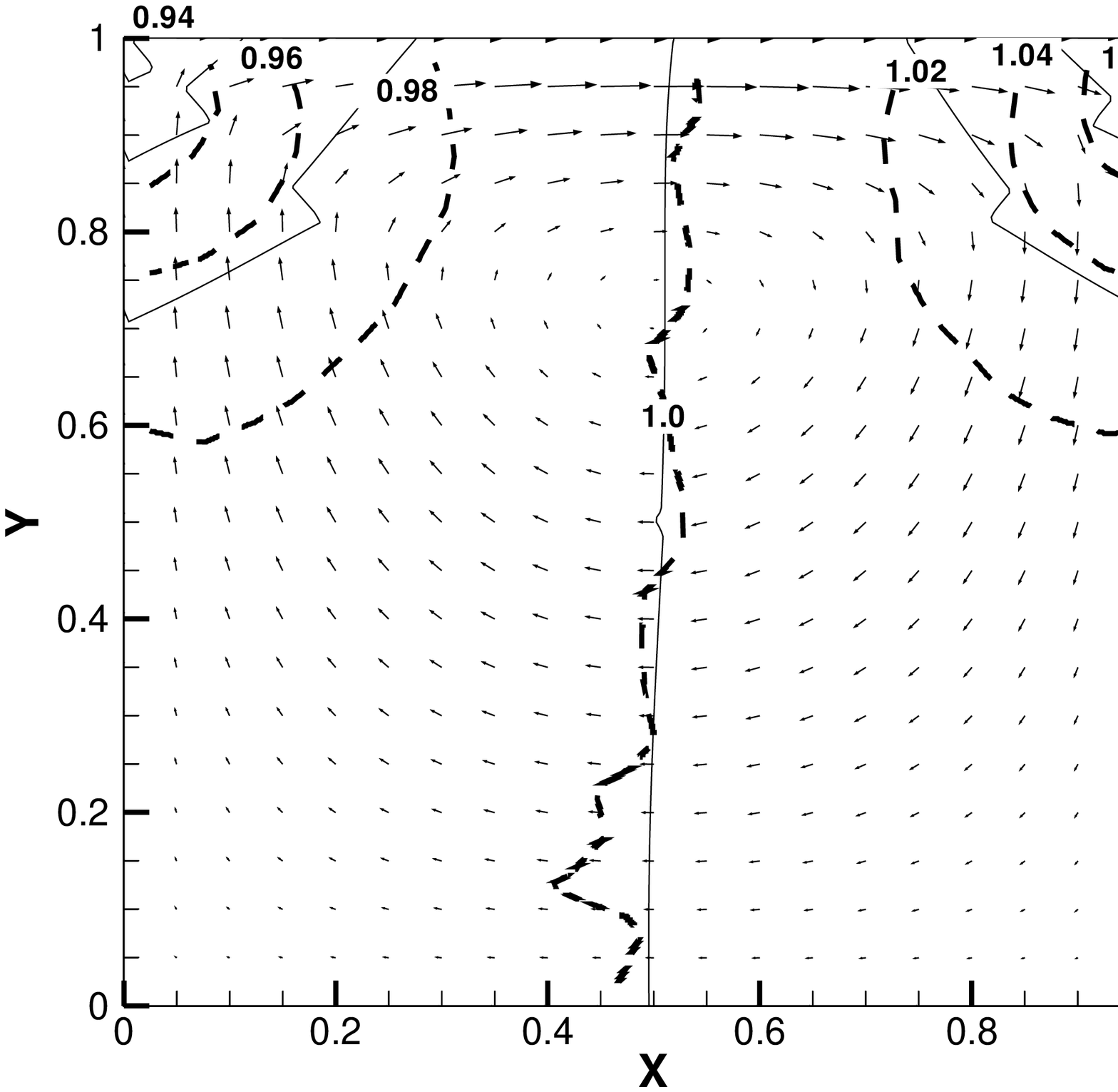}} }
\end{center}
  \caption{Flow in a micro-cavity for $\Kn=0.1$ and ${\rm Ma}=0.14$:
  DSMC simulation \cite[]{RGD} (left) ,
  velocity vector plot and density isolines from ELBM (solid lines)
  with the DSMC density isolines (dashed lines) superimposed (right).}
  \label{dsmc}
\end{figure}

In the hydrodynamic regime, the model was validated using results
available from continuum simulations \cite[]{AK2}. For higher $\Kn
\sim 0.1$, we compared our results with the DSMC simulation of
\cite[]{RGD}. Good agreement between the DSMC simulation and the
ELBM results can be seen in Fig.\ \ref{dsmc}. It can be concluded,
that even for finite Knudsen number, the present model provides
semi-quantitative agreement, as far as the flow profile is
concerned. We remind here again, that the dimensionless density in
the present model corresponds to the dimensionless pressure of a
real fluid so that, for quantitative comparison, the density of
ELBM model should be compared with the pressure computed from
DSMC.

\subsection{Parametric study of the flow in the micro-cavity}

\begin{figure*}
\begin{center}
  \subfigure[]{\includegraphics[width=4cm]{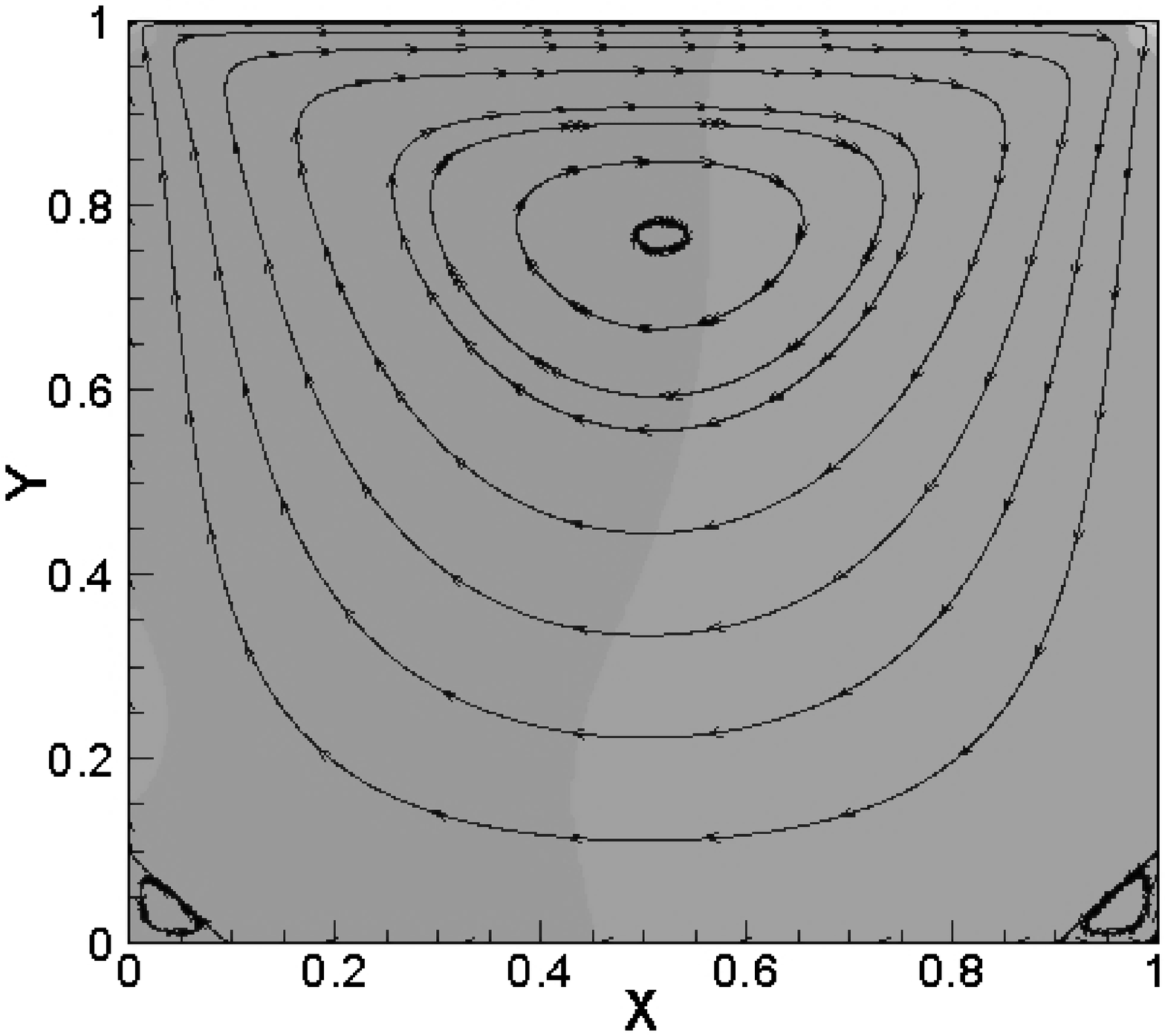}}
  \subfigure[]{\includegraphics[width=4cm]{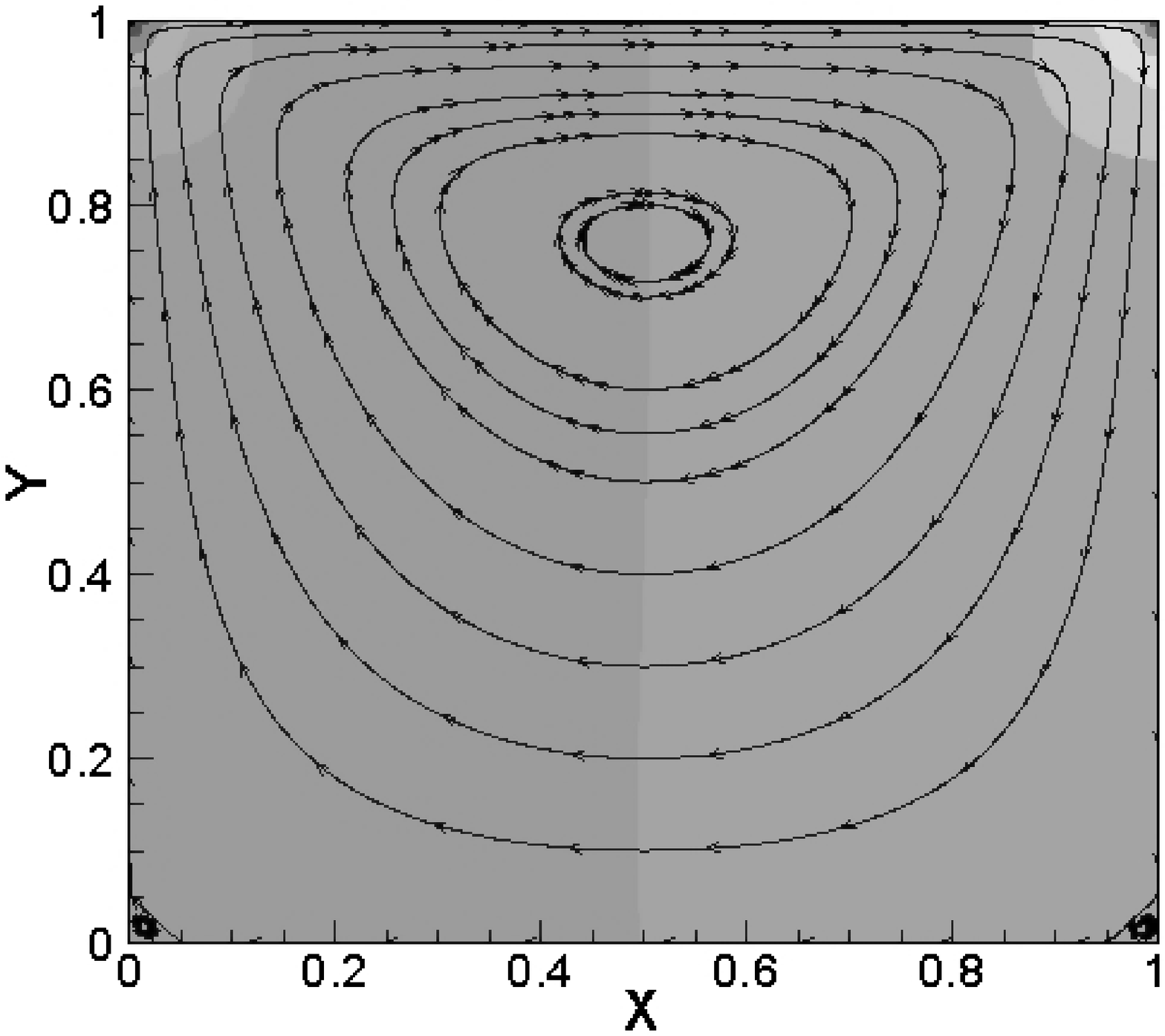}}
  \subfigure[]{\includegraphics[width=4cm]{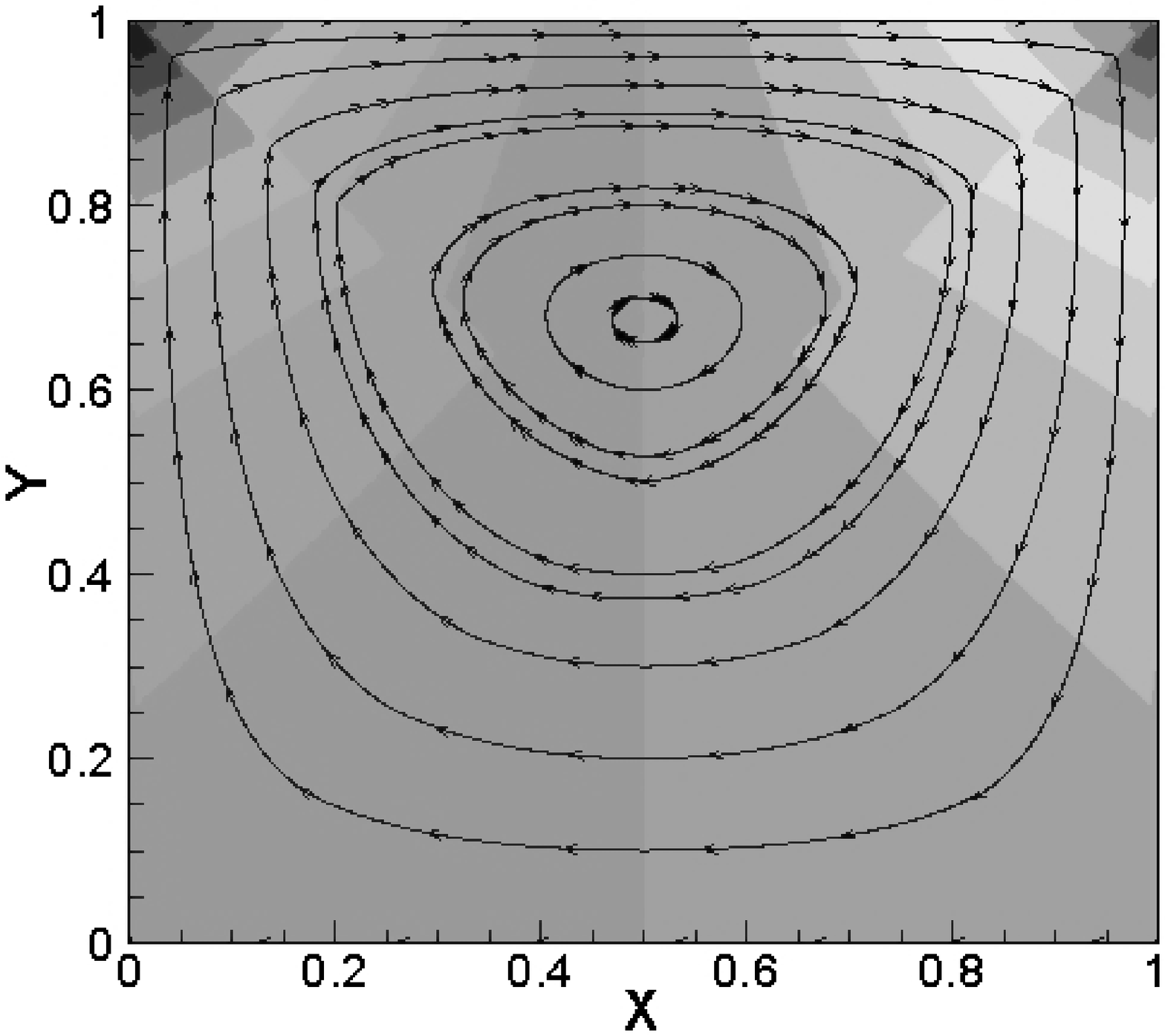}}
\end{center}
\caption{Density isocontours for (a) $\Kn=0.001$, (b)$\Kn=0.01$ ,
and (c) $\Kn=0.1$ (the variation of the density is $0.995\le \rho
\le 1.005$). Superimposed are the streamlines.} \label{rho}
\end{figure*}

Fig.\ \ref{rho} shows the dimensionless density profiles with the
streamlines superimposed for $\Kn=0.001, 0.01,  0.1$. For
$\Kn=0.001$ (${\rm Re}=10$), the behavior expected from continuum
simulations with a large central vortex and two smaller
recirculation zones close to the lower corners can be observed. As
the Knudsen number is increased, the lower corner vortices shrink
and eventually disappear and the streamlines tend to align
themselves with the walls.

The density profiles, as a function of $\Kn$, demonstrate that the
assumption of incompressibility is well justified only in the
continuum regime, where the density is essentially constant away
from the corners. This observation is consistent with the
conjecture that incompressibility requires smallness of the Mach
as well as of the Knudsen number. In hydrodynamic theory, the
density waves decay exponentially fast (with the rate of
relaxation proportional to $\rm \Kn$) leading effectively to
incompressibility. Thus, it is expected that the onset of
incompressibility will be delayed as the Knudsen number increases.

\section{\label{secV} Reduced description of the flow}

The data from the direct simulation of the present kinetic model
were used to validate the effectiveness of different closure
approximations of kinetic theory in the presence of kinetic
boundary layers in a fairly non-trivial flow, and to gain some
insight about the required modification of the closure
approximations in the presence of boundary layers. In this
section, we will present such an analysis for two widely used
closure methods, the Navier-Stokes approximation of the
Chapman-Enskog expansion and Grad's moment closure.

\subsection{The Navier-Stokes approximation}

The Chapman-Enskog analysis \cite[]{ChapK} of the model kinetic
equation leads to a closure relation for the non-equilibrium part
of the pressure tensor as (the Navier-Stokes approximation):
\begin{equation}
\label{NS} \sigma_{xy} = -{\tau} c_{\rm s}^2(\partial_y j_x +
\partial_x j_y).
\end{equation}

\begin{figure*}
\centerline{\includegraphics[width=10cm]{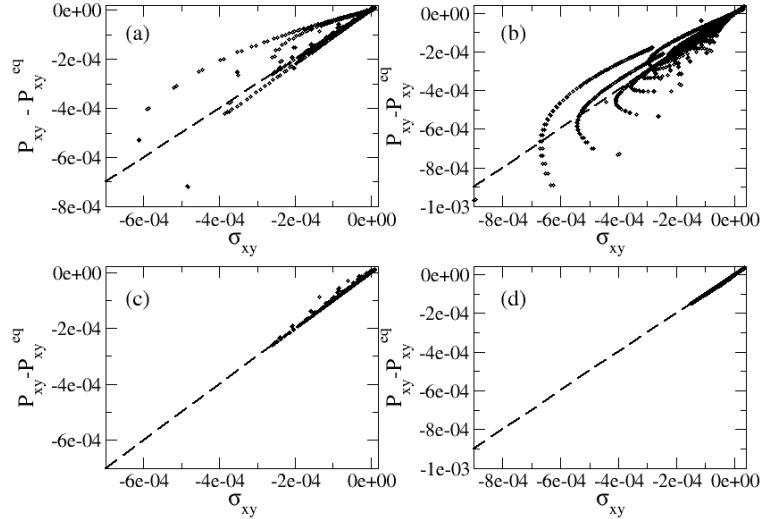}}
\caption{Scatter plot of the non-equilibrium part of the
off-diagonal component of the pressure tensor $P_{xy}-P_{xy}^{\rm
eq}$ and corresponding value computed from Navier-Stokes
approximation $\sigma_{xy}$ (\ref{NS}) for all points in the
domain ((a) and (b)), and after the removal of the boundary layer
corresponding to approximately $10$ mean-free path ((c) and (d)).
Fig.\ (a,c) correspond to $\Kn=0.001$, while Fig.\ (b,d)
correspond to $\Kn=0.01$. Navier-Stokes behavior is indicated by
the straight line of slope equal to one. \label{M4}}
\end{figure*}

Fig.\ \ref{M4} shows a scatter plot of the $xy$ component of the
non-equilibrium part of the pressure tensor  $P_{xy}-P_{xy}^{\rm
eq}$, versus that computed from the Navier-Stokes approximation of
the Chapman-Enskog expansion (\ref{NS}). The upper row is the
scatter plot for all points in the computational domain, while the
lower row is the scatter plot obtained after removal of the
boundary layers close to the four walls of the cavity,
corresponding to approximately $10$ mean-free paths. In all plots,
the dashed straight line of slope equal to one corresponds to
Navier-Stokes behavior. These plots clearly reveal that the
Navier-Stokes description is valid away from the walls in the
continuum as well as in the slip flow regime. On the other hand,
it fails to represent hydrodynamics in the kinetic boundary layer.

Perhaps the most interesting observation from Fig.\ \ref{M4} is
the coherent, curve-like structure of the off-Navier-Stokes
points. These trajectories tend to the Navier-Stokes approximation
as to an attractive sub-manifold. This structure resembles the
trajectories of solutions to the invariance equation
\cite[]{GKbook,GKZ2004} observed, in particular, in a similar
problem of derivation of hydrodynamics from Grad's systems (see,
e. g. \cite{KG203}, p. 831, Fig. 12). A link between solutions to
invariance equations and the present simulations is an intriguing
subject for  further study. In the next subsection, we shall
explore Grad's closure approximation for the present flow.

\subsection{Grad's approximation}

\begin{figure*}
\centerline{\includegraphics[height=4.0cm]{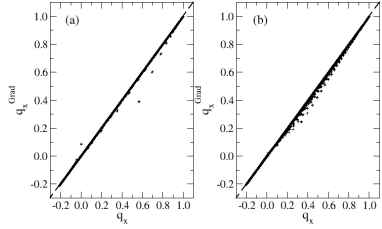}}
\caption{Scatter plot of the computed energy flux, $q_x$, versus
Grad's closure, $q_x^{ \rm Grad}$: (a) $\Kn=0.001$, (b)
$\Kn=0.01$. \label{qGrad}}
\end{figure*}

In contrast to the Chapman-Enskog method,
 the Grad method has an advantage that the approximations are local in space,
 albeit with an increased number of fields.
As the analysis of section \ref{secIII} suggests, the dynamics of
the density, momentum and pressure tensor are almost decoupled
from the rest of the moments, at least away from boundaries. This
motivates the Grad-like approximation for the populations,

\begin{align}
\label{TQA}
\begin{split}
  f_i^{\rm Grad} =W_i\left[ \rho  + \frac{j_{\alpha} c_{i\, \alpha}}{c_{\rm s}^2}+  \frac{1}{ 2  \, c_{\rm s}^4} \,
  \left( P_{\alpha \beta } - \delta_{\alpha \beta } \rho c_{\rm s}^2 \right)
    \left(c_{i\, \alpha} c_{i\, \beta} -  c_{\rm s}^2  \delta_{\alpha \beta }\right)
   \right].
\end{split}
\end{align}
The set of populations parameterized by the values of the density,
momentum and pressure tensor (\ref{TQA}) is a sub-manifold in the
phase space of the system (\ref{LBM}), and can be derived in a
standard way using quasi-equilibrium procedures
\cite[]{GKbook,GKZ2004}. After taking into account the time
discretization, we find the closure relation for the energy flux:
\begin{align}
\label{Gflux}
\begin{split}
q_{\alpha}^{\rm Grad} = \frac{4}{3} \left( 1 + \frac{p}{2 \mu}
\right) {j_{\alpha} }-\frac{p}{2 \mu} q_{\alpha}^{\rm eq}.
\end{split}
\end{align}
In Fig.\ \ref{qGrad}, the scatter plot of the computed energy flux
$q_x$ and the discrete Grad's closure $q_x^{ \rm Grad}$
(\ref{Gflux}) is presented. Same as in Fig.\ \ref{M4}, the
off-closure points in Fig.\ \ref{qGrad} are associated with the
boundary layers. The comparison of the quality with which the
closure relations are fulfilled in Fig.\ \ref{M4} and Fig.\
\ref{qGrad}  clearly indicates the advantage of that a Grad's
closure.
Various strategies of a domain decomposition for a
reduced simulation can be devised based on this observation.
 A
general conclusion is that for slow flows Grad's closure in the
bulk along with the discretization of the boundary condition
\cite[]{AK4} is the optimal strategy for the simulation of
microflows.

\section{\label{secVI}Conclusions and further studies}

We considered a specific example of a minimal kinetic model for
studies of microflows, and compared it with other theories of
nonequilibrium thermodynamics in a nontrivial flow situation. The
close relationship between Grad's moment systems and minimal
kinetic models was highlighted. For the case of a driven cavity
flow, different closure approximations were tested against the
direct simulation data, clearly showing the failure of the
closures near the boundaries. Grad's closure for the minimal model
was found to perform better than the Navier-Stokes approximation.
This finding can be used to reduce the memory requirement in
simulations, while preserving the advantage of locality.

\begin{figure*}
  \begin{center}
  \includegraphics[height=4.0cm]{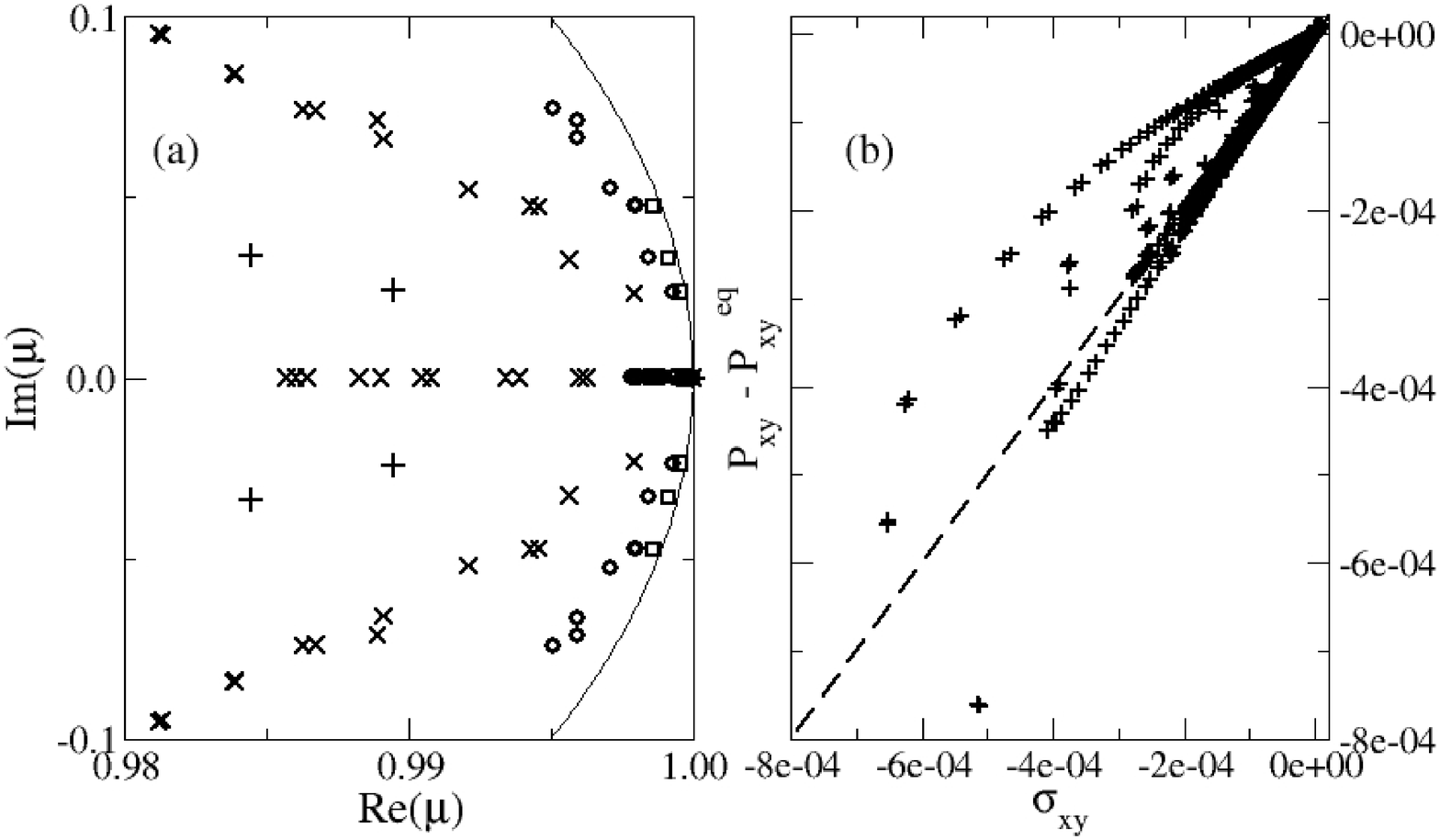}
  \end{center}
  \caption{(a) Leading eigenvalues of the minimal kinetic model at steady state
   (square: Kn=$10^{-4}$, circle: Kn=$10^{-3}$, X: Kn=$10^{-2}$, +: Kn=$10^{-1}$);
   (b) Scatter plot as in Fig.~\ref{M4}(a) for a state perturbed away from the steady
   state along the leading eigenvector (Kn=$10^{-3}$). }
   \label{evals}
\end{figure*}

In this paper, we have explored two classical closures of kinetic
theory. In the future we are going to consider other closures such
as the invariance correction to Grad's closure, and especially
closures based on spectral decomposition \cite[]{GKbook}. To that
end, the ELBM code was coupled with ARPACK \cite[]{arpack} in
order to compute the leading eigenvalues and the corresponding
eigenvectors of the Jacobian field of the corresponding map at the
steady state. In all cases, the eigenvalues are within the unit
circle (Fig.\ \ref{evals}(a)). The leading eigenvalue is always
equal to one (reflecting  mass conservation), and the
corresponding eigenvector captures most of the structure of the
steady state. As the Knudsen number decreases, eigenvalues tend to
get clustered close to the unit circle. This happens because when
the Knudsen number is small the incompressibility assumption is a
good approximation, and mass is also conserved locally. The very
close similarity between fig. \ref{evals}(b) and fig.\
\ref{M4}(a), reveals that states perturbed away from the steady
state along the leading eigenvector are also described well by the
Navier-Stokes closure.

{\it Acknowledgement.} Discussions with A. N. Gorban are
gratefully acknowledged. The work of SA, CEF and IVK was partially
supported by the Swiss Federal Department of Energy (BFE) under
the project Nr. 100862 ``Lattice Boltzmann simulations for
chemically reactive systems in a micrometer domain". The work of
IGK was partially supported by an NSF-ITR grant and by DOE.

\bibliography{cavity}

\begin{thebibliography}{25}
\expandafter\ifx\csname natexlab\endcsname\relax\def\natexlab#1{#1}\fi

\bibitem[Ansumali \& Karlin(2002{\natexlab{{\em a\/}}})]{AK2}
{\sc Ansumali, S. \& Karlin, I.~V.} 2002{\natexlab{{\em a\/}}} Entropy
  {F}unction {A}pproach to the {L}attice {B}oltzmann {M}ethod. {\em J.\ Stat.\
  Phys.\/} {\bf 107(1-2)}, 291--308.

\bibitem[Ansumali \& Karlin(2002{\natexlab{{\em b\/}}})]{AK4}
{\sc Ansumali, S. \& Karlin, I.~V.} 2002{\natexlab{{\em b\/}}} {K}inetic
  {B}oundary {C}ondition for the {L}attice {B}oltzmann {M}ethod. {\em Phys.\
  Rev.\ E\/} {\bf 66(2)}, 026311.

\bibitem[Ansumali {\em et~al.\/}(2004)Ansumali, Karlin, Frouzakis \&
  Boulouchos]{AK2004a}
{\sc Ansumali, S., Karlin, I.~V., Frouzakis, C.~E. \& Boulouchos, K.~B.} 2004
  Entropic {L}attice {B}oltzmann {M}ethod for {M}icroflows. {\em
  http://xxx.lanl.gov/abs/cond-mat/0412555\/} .

\bibitem[Ansumali {\em et~al.\/}(2003)Ansumali, Karlin \& \"Ottinger]{AK5}
{\sc Ansumali, S., Karlin, I.~V. \& \"Ottinger, H.~C.} 2003 Minimal {E}ntropic
  {K}inetic {M}odels for {S}imulating {H}ydrodynamics. {\em Europhys. Lett.\/}
  {\bf 63(6)}, 798--804.

\bibitem[Beskok \& Karniadakis(2001)]{Karniadakis2}
{\sc Beskok, A. \& Karniadakis, G.~E.} 2001 {\em Microflows: {F}undamentals and
  {S}imulation\/}. Springer, Berlin.

\bibitem[Bird(1994)]{bird}
{\sc Bird, G.~A.} 1994 {\em Molecular {G}as {D}ynamics and the {D}irect
  {S}imulation of {G}as {F}lows. {T}heory and {A}pplication of the {B}oltzmann
  {E}quation\/}. Oxford University Press.

\bibitem[Cercignani(1975)]{Cerci}
{\sc Cercignani, C.} 1975 {\em Theory and {A}pplication of the {B}oltzmann
  {E}quation\/}. Scottish Academic Press, Edinburgh.

\bibitem[Chapman \& Cowling(1970)]{ChapK}
{\sc Chapman, S. \& Cowling, T.~G.} 1970 {\em The {M}athematical {T}heory of
  {N}on-{U}niform {G}ases\/}. Cambridge University Press, Cambridge.

\bibitem[Gorban \& Karlin(1996)]{GK96a}
{\sc Gorban, A.~N. \& Karlin, I.~V.} 1996 Short-wave {L}imit of
  {H}ydrodynamics: {A} {S}oluble {E}xample. {\em Phys. Rev. Lett.\/} {\bf 77},
  282--285.

\bibitem[Gorban \& Karlin(2005)]{GKbook}
{\sc Gorban, A.~N. \& Karlin, I.~V.} 2005 {\em Invariant Manifolds for Physical
  and Chemical Kinetics\/}. Springer, Berlin Heidelberg.

\bibitem[Gorban {\em et~al.\/}(2004)Gorban, Karlin \& Zinovyev]{GKZ2004}
{\sc Gorban, A.~N., Karlin, I.~V. \& Zinovyev, A.~Y.} 2004 {C}onstructive
  {M}ethods of {I}nvariant {M}anifolds for {K}inetic {P}roblems. {\em Phys.
  Rep.\/} {\bf 396}, 197--403.

\bibitem[Grad(1949)]{GradH}
{\sc Grad, H.} 1949 On the {K}inetic {T}heory of {R}arefied {G}ases. {\em
  Comm.\ Pure Appl.\ Math.\/} {\bf 2}, 331--407.

\bibitem[Jiang {\em et~al.\/}(2003)Jiang, J. \& Shen]{RGD}
{\sc Jiang, J.-Z., J., F. \& Shen, C.} 2003 Statistical simulation of
  micro-cavity flows. {\em 23rd Int. Symposium on Rarefied Gas Dynamics\/} pp.
  784--790.

\bibitem[Karlin {\em et~al.\/}(1999)Karlin, Ferrante \& \"Ottinger]{DHT}
{\sc Karlin, I.~V., Ferrante, A. \& \"Ottinger, H.~C.} 1999 Perfect {E}ntropy
  {F}unctions of the {L}attice {B}oltzmann {M}ethod. {\em Europhys.\ Lett.\/}
  {\bf 47}, 182--188.

\bibitem[Karlin \& Gorban(2002)]{KG203}
{\sc Karlin, I.~V. \& Gorban, A.~N.} 2002 Hydrodynamics from {G}rad's
  {E}quations: {W}hat can {W}e {L}earn from {E}xact {S}olutions? {\em Ann.\
  Phys. (Leipzig)\/} {\bf 11}, 783--833.

\bibitem[Kevrekidis {\em et~al.\/}(2003)Kevrekidis, Gear, Hyman, Kevrekidis,
  Runborg \& Theodoropoulos]{Yanis}
{\sc Kevrekidis, I.~G., Gear, C., Hyman, J.~M., Kevrekidis, P.~G., Runborg, O.
  \& Theodoropoulos, C.} 2003 {E}quation-free, {C}oarse-grained {M}ultiscale
  {C}omputation: {E}nabling {M}icroscopic {S}imulators to {P}erform {S}ystem
  -level {A}nalysis. {\em Comm. Math. Sci.\/} {\bf 1}, 715--762.

\bibitem[Kevrekidis {\em et~al.\/}(2004)Kevrekidis, Gear \& Hummer]{Yannis4}
{\sc Kevrekidis, I.~G., Gear, C.~W. \& Hummer, G.} 2004 Equation-free: the
  {C}omputer-assisted {A}nalysis of {C}omplex, {M}ultiscale systems. {\em A. I.
  Ch. E. Journal\/} {\bf 50}~(7), 1346--1354.

\bibitem[Lehoucq {\em et~al.\/}(1998)Lehoucq, Sorensen \& Yang]{arpack}
{\sc Lehoucq, R., Sorensen, D. \& Yang, C.} 1998 {\em {ARPACK} Users' Guide:
  Solution of Large-Scale Eigenvalue Problems with Implicitly Restarted Arnoldi
  Methods\/}. SIAM.

\bibitem[Niu {\em et~al.\/}(2003)Niu, Shu \& Chew]{ELBMMICRO}
{\sc Niu, X.~D., Shu, C. \& Chew, Y.} 2003 Lattice {B}oltzmann {BGK} {M}odel
  for {S}imulation of {M}icro {F}lows. {\em Euro. Phys. Lett.\/} {\bf 67},
  600--606.

\bibitem[Oran {\em et~al.\/}(1998)Oran, Oh \& Cybyk]{DSMC}
{\sc Oran, E.~S., Oh, C.~K. \& Cybyk, B.~Z.} 1998 {D}irect {S}imulation {M}onte
  {C}arlo: {R}ecent {A}dvances and {A}pplications. {\em Annu\ Rev.\ Fluid
  Mech.\/} {\bf 30}, 403--441.

\bibitem[Slemrod(1998)]{Slemrod98}
{\sc Slemrod, M.} 1998 Renormalization of the {C}hapman-{E}nskog {E}xpansion:
  {I}sothermal {F}luid {F}low and {R}osenau {S}aturation. {\em J. Stat.
  Phys.\/} {\bf 91}, 285--305.

\bibitem[Succi \& Sbragaglia(2004)]{SUCCIDBC}
{\sc Succi, S. \& Sbragaglia, M.} 2004 Analytical {C}alculation of {S}lip
  {F}low in {L}attice {B}oltzmann {M}odels with {K}inetic {B}oundary
  {C}onditions. {\em http://arxiv.org/abs/nlin.CG/0410039\/} .

\bibitem[Theodoropoulos {\em et~al.\/}(2000)Theodoropoulos, Qian \&
  Kevrekidis]{Yannis3}
{\sc Theodoropoulos, K., Qian, Y.-H. \& Kevrekidis, I.~G.} 2000 ``{C}oarse"
  {S}tability and {B}ifurcation {A}nalysis {U}sing {T}imesteppers: a reaction
  diffusion example. {\em Proc. Natl. Acad. Sci.\/} {\bf 97}~(18), 9840--9843.

\bibitem[Theodoropoulos {\em et~al.\/}(2004)Theodoropoulos, Sankaranarayanan,
  Sundaresan \& Kevrekidis]{Yanis1}
{\sc Theodoropoulos, K., Sankaranarayanan, K., Sundaresan, S. \& Kevrekidis,
  I.~G.} 2004 Coarse {B}ifurcation {S}tudies of {B}ubble {F}low {L}attice
  {B}oltzmann {S}imulations. {\em Chem. Eng. Sci.\/} {\bf 59}, 2357--2362.

\bibitem[Zheng {\em et~al.\/}(2002)Zheng, Garcia \& Alder]{Garcia}
{\sc Zheng, Y., Garcia, A.~L. \& Alder, B.~J.} 2002 {C}omparison of {K}inetic
  {T}heory and {H}ydrodynamics for {P}oiseuille {F}low. {\em J. Stat. Phys.\/}
  {\bf 109}, 495--505.

\end{thebibliography}
\bibliographystyle{jfm}

\end{document}